\pdfoutput=1
\RequirePackage[T1]{fontenc}
\documentclass[12pt]{article}

\usepackage[height=8.85in,width=6.45in]{geometry}

\usepackage[whole,scale=.9]{bxcjkjatype} 

\usepackage[utf8]{inputenc}
\usepackage{amsmath}
\usepackage{amssymb}
\usepackage{mathtools}
\numberwithin{equation}{section}
\usepackage{slashed}
\usepackage{braket}
\usepackage{enumitem}
\usepackage[svgnames]{xcolor}
\usepackage[colorlinks,citecolor=DarkGreen,linkcolor=FireBrick,urlcolor=FireBrick,linktocpage,unicode,psdextra]{hyperref}
\usepackage{cite}
\usepackage{graphicx}
\usepackage{tikz}
\usepackage{tikz-cd}

\tikzset{>=stealth}
\usepackage[bottom]{footmisc}

\usepackage{times}
\usepackage{courier}
\usepackage{bm}
\usepackage{subfig}

\usepackage{xcolor}
\usepackage{mdframed}

\renewenvironment{figure}[1][]{
  \begin{originalfigure}[#1]
    \begin{mdframed}[linecolor=black!0,backgroundcolor=black!1]
}{
    \end{mdframed}
  \end{originalfigure}
}

\usepackage{ascmac}
\usepackage{dashbox}
\usepackage{xcolor}
\usepackage{colortbl}
\usepackage{arydshln}
\definecolor{lightyellow}{rgb}{1.0, 0.95, 0.7}
\definecolor{blue}{rgb}{0.0, 0.4, 1.0}
\definecolor{Blue}{rgb}{0,0,1}
\definecolor{darkgreen}{rgb}{0.,0.6,0.}

\definecolor{colorA}{rgb}{1,0,0}
\definecolor{colorB}{rgb}{0,0.3,1}
\definecolor{colorC}{rgb}{0.9,0.8,0.2}
\definecolor{colorD}{rgb}{0,0.65,0}
\definecolor{lesslightgray}{rgb}{0.5,0.5,0.5}

\let\oldmathcal\mathcal
\let\mathcal\mathsf

\def\Nequals#1{$\oldmathcal{N}{=}#1$}

\begin{document}

\begin{titlepage}

\begin{flushright}
\end{flushright}

\vskip 3cm

\begin{center}

{\Large \bfseries Undecidable problems in quantum field theory}

\vskip 1cm
Yuji Tachikawa
\vskip 1cm

\begin{tabular}{ll}
 & Kavli Institute for the Physics and Mathematics of the Universe (WPI), \\
& University of Tokyo,  Kashiwa, Chiba 277-8583, Japan
\end{tabular}

\vskip 2cm

\end{center}

\noindent 
We point out that some questions in quantum field theory are undecidable in a precise mathematical sense. 
More concretely, it will be demonstrated that there is no algorithm answering whether a given 2d supersymmetric Lagrangian theory breaks supersymmetry or not.
It will also be shown that there is a specific 2d supersymmetric Lagrangian theory which breaks supersymmetry if and only if the standard Zermelo-Fraenkel set theory with the axiom of choice is consistent,
which can never be proved or disproved as the consequence of G\"odel's second incompleteness theorem.
The article includes a brief and informal introduction to the phenomenon of undecidability and its previous appearances in theoretical physics.

\bigskip

A 15-minute video presentation of the content is also available at \url{https://www.youtube.com/watch?v=H548i3dnsWE}.

\end{titlepage}

\setcounter{tocdepth}{2}
\tableofcontents


\section{Introduction}
\label{sec:introduction}

Many aspects of life are often unpredictable,
and physics is no exception.
By now, physicists are familiar with two types of unpredictability:
one is the inherent probabilistic nature of quantum mechanics,
and another is the chaotic behavior of various classical systems,
where a minute difference in the initial condition is exponentially amplified,
leading to the loss of predictability. 
In this article, we would like to introduce the audience to a third type of unpredictability,
whose existence does not seem to be too widely appreciated. 

To set up the stage, consider physicists working on a class of systems $X$.
They are often interested in a certain property $P$ of those systems,
and they often spend their entire academic career
trying to answer the following question: \begin{quotation}
Given a system $x\in X$ in this class, does it have the property $P$?
\end{quotation}
Examples of such $(X,P)$ are:
\begin{center}
\begin{tabular}{|c|c|c|}
\hline
&$X$ & $P(x)$ for $x\in X$ \\
\hline\hline
1)&quantum spin chains & $x$ is gapless in the infinite volume limit \\
\hline
2)&supersymmetric theories & $x$ spontaneously breaks supersymmetry \\
\hline
3)&condensed matter systems & $x$ is superconducting at room temperature and pressure\\
\hline
4)&gauge theories & $x$ confines \\
\hline
\end{tabular}
\end{center}
The  unpredictability which we will discuss comes in two types, one generic and another specific.
The generic one  is of the following form:
\begin{quote}
There can be no single algorithm which decides for every $x\in X$ whether $x$ has the property $P$,
\end{quote} whereas the specific one has the form: \begin{quote}
There is a concrete $x\in X$ for which it is not possible to prove or disprove whether $x$ has the property $P$.
\end{quote} 
As we will see below, undecidability in both senses would usually arise together,
so we will simply call such a problem $(X,P)$ undecidable.

Such situations are well-known in  the theory of computation and in mathematical logic.
There is also a long history of works showing that there are similarly undecidable questions in theoretical physics.\footnote{%
That said, the author of this article was totally unaware of such series of works, and naively thought, before Beni Yoshida introduced him to \cite{Cubitt2nd} in May 2021, that whether 1d spin chains are gapless
and whether 2d supersymmetric Lagrangian theories break supersymmetry
would be algorithmically decidable in the future when the theoretical physics would be sufficiently developed.
The author wanted to share his surprise with his colleagues and gave an informal lunch talk on this topic in autumn 2021.
The reaction of mathematicians there was,
however, that the author of the present article was too naive and was behind mathematicians by about 100 years.
Therefore it is perfectly plausible that many physicists reading this article might similarly laugh at the na\"\i v\'et\'e of the author.
But the author still wanted to convey his surprise to a broader community, which was the main motivation for the author to prepare this article.
It should also be mentioned that there are many statements in mathematics which are undecidable in the same sense, even though the statements themselves might not seem to have a connection to mathematical logic at a first glance.
For a nice exposition of such statements, the readers are referred to \cite{Poonen}.
}
The earliest properly formulated statement of this kind in theoretical physics that the author of the present article is aware of is\footnote{%
The author based this historical information on the introduction of an article \cite{Enc} suggested by one of the referees.
(The main physics claim of \cite{Enc}, presented in its Sec.~2, was too sketchy for the author of this present article to judge its validity. But the introduction does contain valuable encyclopedic information.)
The same referee also suggested the author of the present article to cite \cite{BY,Epi}.
The former article \cite{BY} is an essay on the relation of G\"odel's incompleteness theorem and physics,
the validity of which is beyond the ability of the author of the present article to decide.
The latter article \cite{Epi} is a philosophical analysis of the series of works starting in \cite{2dNature}.
As the author of the present article is not well-versed in the philosophy of science,
it is again beyond the capability of the author of the present article to discuss the validity of the content,
but the author of the present article still wants to mention that two out of the three main points listed under bullet points in their summary section contain  factual errors in the interpretation of \cite{2dNature} and quantum spin systems in general.
Namely, the first  claims that the local form of the Hamiltonian of \cite{2dNature} depends on the size of the lattice, which is not the case.
Similarly, the second of the three claims that any finite quantum spin systems can be numerically computable on an actual computer, thus making the undecidability claim of \cite{2dNature} meaningless.
But an actual computation of a large quantum spin system is exponentially hard due to the exponential growth of the number of states. As such, any such philosophical analysis needs to tackle directly the question of the meaning of philosophical computability of a computation practically impossible, which was not adequately addressed in that article.
} the work \cite{GerochHartle}, where the question \begin{equation}
\begin{aligned}
X&=\{\text{Classical mechanical systems}\},\\
P(x)&=\text{$x$ leads to a specified outcome}
\end{aligned}
\end{equation}
was shown to be undecidable. 
Shortly afterwords, works \cite{Moore,daCostaDoria} also discussed undecidable questions in classical mechanics.
More recently, it was shown in \cite{2dNature,2dLong} from 2015, that the question \begin{equation}
\begin{aligned}
X&=\{\text{2d quantum spin systems}\},\\
P(x)&=\text{$x$ is gapless in the infinite volume limit}
\end{aligned}
\end{equation}
 was similarly demonstrated to be undecidable. 
This result was later strengthened to the undecidability of 
gaplessness of 1d quantum spin systems in  \cite{1d,Cubitt2nd}
and extended to the undecidability of the phase space of 2d quantum spin systems
and of the outcome of the real-space renormalization group flow 
in \cite{PhaseDiagram,RG}, respectively.
Similarly, in \cite{Control}, it was shown that whether a quantum optimal control protocol can be implemented or not is undecidable,
and in the work \cite{thermalNature,thermalLong}, it was demonstrated that 
whether 1d quantum spin systems thermalize is undecidable.
In the present article, we will also see that the question \begin{equation}
\begin{aligned}
X&=\{\text{2d \Nequals{(2,2)} supersymmetric Lagrangian theories}\},\\
P(x)&=\text{$x$ breaks supersymmetry}
\end{aligned}
\end{equation} is equally undecidable. 

In each case, the undecidability is shown by reducing it to the most basic and fundamental case of undecidability, the halting problem of computer programs: \begin{equation}
\begin{aligned}
\Xi & =\{ \text{computer programs} \},\\
\Pi(\xi) & = \text{the execution of $\xi$ halts in finite time.}
\end{aligned}
\end{equation}
As will be recalled later, there can be no algorithm to decide $\Pi(\xi)$ for each $\xi\in \Xi$.
Furthermore, one can write a program $\xi_0$ which halts in finite time
if  and only if the Zermelo-Fraenkel set theory with the axiom of choice,
the standard foundation of mathematics of our times, is inconsistent.
Due to  G\"odel's second incompleteness theorem, that $\xi_0$ halts in finite time can neither be proved nor disproved. 
Therefore, the remaining task to establish the undecidability of a particular question $(X,P)$ 
is to construct a concrete map $f:\Xi\to X$ such that $P(f(\xi))=\Pi(\xi)$,
i.e.~to find a way to construct systems $f(\xi)\in X$ encoding computer programs $\xi$ such that $f(\xi)$ has the property $P$ if and only if the program $\xi$ halts in finite time.

The rest of the article is organized as follows.
In Sec.~\ref{sec:preliminaries}, we collect basic facts concerning undecidability in computer science and mathematical logic.
In more detail, we very briefly review the halting problem of Turing machines in Sec.~\ref{sec:halting},
summarize G\"odel's incompleteness theorems in Sec.~\ref{sec:incompleteness},
discuss what their combination means in Sec.~\ref{sec:combination}.
We then recall in Sec.~\ref{sec:diophantine} their relation to the negative solution to Hilbert's 10th problem,
i.e.~the fact that the solvability of Diophantine equations is undecidable.

We then come to Sec.~\ref{sec:physics}, which contains three physics manifestations of undecidability.
In Sec.~\ref{sec:classical}, we recall the arguments of \cite{GerochHartle,Moore,daCostaDoria} who found the undecidability in classical mechanics.
In Sec.~\ref{sec:gaplessness}, we briefly review the results of \cite{2dNature,2dLong},
which showed the undecidability of the gaplessness of 2d spin systems.
Then in Sec.~\ref{sec:susy}, we show that whether 2d supersymmetric Lagrangian theory breaks supersymmetry is undecidable, 
which is a simple consequence of the negative solution to Hilbert's 10th problem.
We will conclude with discussions in Sec.~\ref{sec:discussions}.

\section{Preliminaries}
\label{sec:preliminaries}
In this section, we give a brief and informal overview of fundamental results we need from theoretical computer science and mathematical logic.
Interested readers should consult standard textbooks for more details.
For those who read Japanese, an extremely readable account for non experts is \cite{Terui}.

\subsection{Halting problem}
\label{sec:halting}
Computers are ubiquitous in modern life, but the theory of computing predates modern digital computers and goes back to the early 20th century.
At that time, multiple proposals were made to formalize precisely what it means to compute algorithmically. 
Various proposals were made, including Turing machines, Church's $\lambda$ calculus, and recursive functions.
These were soon understood  to be all equivalent
since an algorithm expressed in one formulation can be translated to an algorithm expressed in another.
We can then call something computable when it is computable in any one of these equivalent formulations;
this is the Church-Turing thesis.

In the discussion below, we typically consider Turing machines, which are in some sense the closest among the formulations to the actual computers we use. 
A Turing machine $\xi$ has a finite-length program and a finite number of internal states, and works with an infinite amount of memory.
It operates step by step, according to its program, by updating its internal state and modifying a finite amount of memory at each step.

We note that a Turing machine accepting a finite sequence of natural numbers $n_1,n_2,\ldots$ 
can also be thought of as a Turing machine accepting a single natural number.
This is because e.g. we can encode $n_1,n_2,\ldots$ into a single natural number
 $N:=2^{n_1} 3^{n_2} \cdots (p_k)^{n_k}\cdots$ where $p_k$ is a $k$-th prime number,
 and that we can recover $n_1,n_2,\ldots$ from $N$ algorithmically. 
This allows us to be flexible about the number of arguments a Turing machine takes.

Consider a Turing machine $\xi$ accepting a natural number $n$ as the input. 
It then tries to compute the result, emit it as an output $\xi(n)$, and halt.
It can also go into an infinite loop and does not halt,
as often happens when one first learns to program.
In this case the output is undefined.
Below, we simply say $\xi(n)$ halts or does not halt, to distinguish these two behaviors.

We need two fundamental facts of Turing machines.
One is the existence of a universal Turing machine $\upsilon$, which is a single program which emulates arbitrary Turing machines in the following sense.
Note that a program describing a Turing machine $\xi$ is a finite string of alphabets. 
It can be saved in a text file, which is simply a string of bits (zeros and ones). 
It can then be regarded as a natural number $\lceil \xi\rceil$ written in binary digits.
A universal Turing machine $\upsilon$ accepts two natural numbers, $i$ and $n$.
If $i$ is of the form $\lceil\xi\rceil$ for a Turing machine $\xi$, i.e.~if $i$ describes a valid program,
$\upsilon$ emulates the behavior of $\xi$ operating on $n$,
and emits $\xi(n)$ as the output if $\xi$ halts, and $\upsilon$ runs forever if $\xi$ runs forever.
Schematically, we have $\upsilon(\lceil\xi\rceil,n)=\xi(n)$.
The existence of $\upsilon$ is nontrivial, but we do use this fact every day:
the CPU can run arbitrary programs stored in the memory, since it is a universal Turing machine.

Another fundamental fact of Turing machines is that it is impossible to write a Turing machine $\eta$
such that \begin{equation}
\eta(\lceil \xi\rceil,n)=\begin{cases}
0 & \text{if $\xi(n)$ halts},\\
1 & \text{if $\xi(n)$ does not halt}.
\end{cases}
\end{equation} 
In other words, the halting problem of Turing machines is algorithmically undecidable. 
It can be proved as follows. Suppose there is such an $\eta$. 
We can then write a Turing machine $\delta$, which accepts a natural number $i$ as the input
and does the following:
\begin{itemize}
\item Compute $\eta(i,i)$.
\item If the result is $0$, go to an infinite loop.
\item If the result is $1$, halt.
\end{itemize}
Now consider the behavior of $\delta(\lceil\delta\rceil)$. 
Expanding the definition, we see that it  does the following: \begin{itemize}
\item Compute $\eta(\lceil\delta\rceil,\lceil\delta\rceil)$. The result is $0$ if $\delta(\lceil\delta\rceil)$ halts, and $1$ if $\delta(\lceil\delta\rceil)$ does not halt.
\item If the result is $0$, go to an infinite loop.
\item If the result is $1$, halt.
\end{itemize}
Therefore, $\delta(\lceil\delta\rceil)$ halts if $\delta(\lceil\delta\rceil)$ does not halt,
and $\delta(\lceil\delta\rceil)$ does not halt if $\delta(\lceil\delta\rceil)$ halts, which is a contradiction.

\subsection{Incompleteness theorems}
\label{sec:incompleteness}

We now proceed to an informal discussion of  the incompleteness theorems of G\"odel. 
First we need to introduce a few technical terminologies.
A formal system, or equivalently a theory, is a set of axioms and inference rules.
Let us denote such a theory by $\mathcal{T}$.
A model $M$ of $\mathcal{T}$ is a mathematical object satisfying these axioms and rules.
As an example, we can consider a theory $\mathcal{G}$ of groups,
consisting of the usual axioms of groups.
Tautologically, a model of this theory $\mathcal{G}$ is a group.
As another example, we can take the Zermelo-Fraenkel set theory with the axiom of choice,
which is usually abbreviated as $\mathcal{ZFC}$.
This is often taken as the foundation of modern mathematics.
In the  case of $\mathcal{ZFC}$, we do not usually think of using a particular model of it,
but we are effectively working in one. 

Consider a mathematical statement $\phi$ in a theory $\mathcal{T}$.
A proof $\pi$ of $\phi$ in the theory $\mathcal{T}$ is a series of valid inferences starting from the axioms of $\mathcal{T}$.
If there is a proof $\pi$ of $\phi$ in the theory $\mathcal{T}$, 
the statement $\phi$ is true in any model $M$.
This is known as the soundness of logic.
A more nontrivial fact is that if a statement $\phi$ is true in every model $M$,
$\phi$ has a proof in the theory $\mathcal{T}$.
This is known as the completeness of logic.
A theory $\mathcal{T}$ is called consistent if there is no proof of a contradiction in the theory.
This is known to be equivalent for the theory $\mathcal{T}$ to have a model $M$.
We assume below that all theories we discuss  are consistent, unless otherwise mentioned.

For example, we can consider the theory $\mathcal{G}$ of groups, 
and let $\phi_\text{Ab}$ be the statement that all elements commute with each other.
The statement $\phi_\text{Ab}$ is true or false in a model $G$ of $\mathcal{G}$,
depending on whether $G$ is Abelian or not.
Then, the statement $\phi_\text{Ab}$ can neither be proved nor be disproved in the theory $\mathcal{G}$ of groups, 
and whether $\phi_\text{Ab}$ is true or not depends on the particular model $G$ of the theory $\mathcal{G}$.
The combinations $\mathcal{G}+\phi_\text{Ab}$
and 
$\mathcal{G}+\neg\phi_\text{Ab}$ are the theory of Abelian groups
and the theory of non-Abelian groups, respectively.

In general, it should not be surprising 
that a certain statement $\phi$ in a theory $\mathcal{T}$ can neither be proved nor be disproved,
and whether $\phi$ is true or not depends on the chosen model $M$ of $\mathcal{T}$.
Such statement $\phi$ is said to be independent of the  theory $\mathcal{T}$.

In the early 20th century, when people first started to formalize mathematics, 
it was realized that the set theory can be used as a foundation of any other known mathematics,
by encoding other mathematical objects as sets;
it is similar to the use of the assembly language to implement higher-level programming languages.
It was then hoped that the set theory would either prove or disprove any statement $\phi$,
and therefore assign whether each statement $\phi$ is true or false uniquely.

G\"odel's first incompleteness theorem says that this is impossible. 
Namely, it says that for any consistent theory $\mathcal{T}$ capable of describing elementary arithmetic of natural numbers, 
there is a statement $\zeta_{\mathcal{T}}$ of this theory which is independent of $\mathcal{T}$.
As we said, this means that $\zeta_{\mathcal{T}}$ can neither be proved nor disproved in this theory,
and therefore the validity of the statement $\zeta_{\mathcal{T}}$ depends on the model of $\mathcal{T}$.
The proof of this theorem uses the idea that the statements $\phi$ and the proofs $\pi$ in the theory $\mathcal{T}$ 
can be saved in text files, whose content can be considered as (gigantic) natural numbers $\lceil \phi\rceil$ and $\lceil\pi\rceil$.
These are called G\"odel codes of $\phi$ and $\pi$, respectively.
Then, whether a statement $\phi$ can be proved or not can itself be  rephrased as an arithmetical statement 
about its G\"odel code $\lceil\phi\rceil$ which can be treated within the theory $\mathcal{T}$.

To state G\"odel's second incompleteness theorem,
we consider the statement that there is no proof in the theory $\mathcal{T}$ which leads to a contradiction.
This statement can also be translated into an arithmetical statement $\text{Con}(\mathcal{T})$,
which says there is no natural number representing the content of a text file containing a proof of a contradiction.
Then G\"odel's second incompleteness theorem says that $\text{Con}(\mathcal{T})$ can neither be proved nor be disproved in $\mathcal{T}$.
Our discussions have been extremely rough, but even more roughly speaking, this means the following:
if a theory $\mathcal{T}$ is consistent and is capable of describing elementary arithmetic of natural numbers, 
it can never prove or disprove its own consistency. 

These incompleteness theorems apply to $\mathcal{ZFC}$, the Zermelo-Fraenkel set theory with the axiom of choice,
which is usually taken as the standard foundation of modern mathematics. 
This means that $\mathcal{ZFC}$ can never prove or disprove $\text{Con}(\mathcal{ZFC})$.\footnote{%
Analyzing the structure of $\mathcal{ZFC}$ mathematically using $\mathcal{ZFC}$ might sound circular, 
but it is simply a special case of studies of various formal systems $\mathcal{T}$ using $\mathcal{ZFC}$ assuming the consistency of $\mathcal{ZFC}$, 
where $\mathcal{T}$ happens to be $\mathcal{ZFC}$.
As such, we can learn a lot about $\mathcal{T}=\mathcal{ZFC}$, 
for example the fact that it cannot prove its own consistency.

The author thinks that it is vaguely analogous to the situation of the validity of the Born rule and the theory of measurement in quantum mechanics. 
There, one analyzes the combined system of the experimental apparatus and the target system by applying the Born rule to the entire system.
In this manner one can learn a lot about the measurement process, although one cannot prove the validity of the Born rule as applied to the entire system.
}
Adding $\text{Con}(\mathcal{ZFC})$ as an axiom does not improve the situation,
since the consistency of the combination $\mathcal{ZFC}+\text{Con}(\mathcal{ZFC})$ cannot be proved from it either.
A more peculiar point is that the combination $\mathcal{ZFC} + \neg \text{Con}(\mathcal{ZFC})$ is also consistent if $\mathcal{ZFC}$ is consistent.
This somewhat contradictory situation is possible because,
in a model $M$ of $\mathcal{ZFC} + \neg \text{Con}(\mathcal{ZFC})$,
the statement `there is an natural number which encodes the proof of a contradiction' is true in $M$,
but what $M$ thinks as the set of natural number can be larger than what $\mathbb{N}$ is for us.

\subsection{Incompleteness theorems and Turing machines}
\label{sec:combination}
Let us now combine what we learned about Turing machines and incompleteness theorems.
Suppose one wants to know if a statement $\phi$ is provable in a theory $\mathcal{T}$.
Let us write $\phi$ in an editor, save it into a text file, and regard its content as a binary representation of a natural number $\lceil \phi \rceil$.
We can then write  a program $\gamma_{\mathcal{T}}(\lceil\phi\rceil, n)$ which does the following: 
\begin{itemize}
\item Regard the natural number $n$ as the binary representation $\lceil\pi \rceil$ of the content of a text file containing a purported proof of $\phi$.
\item Check whether $\pi$ is a valid proof of $\phi$ in the theory ${\mathcal{T}}$. 
\end{itemize}
We note that such programs $\gamma_{\mathcal{T}}$ are known as proof verifiers and their development is an active area of research.\footnote{%
For example, G\"odel's second incompleteness theorem itself was checked  several years ago in \cite{Paulson1,Paulson2} using the proof verifier Isabelle/HOL.
For informed opinions of pure mathematicians who advocate the use of proof verifiers from their own experiences of wrong proofs, 
see e.g.~\cite{talkVoevodsky,talkBuzzard}.
}

Using $\gamma_{\mathcal{T}}$,
it is possible to write a program $\delta_{\mathcal{T}}(\lceil\phi\rceil)$ 
which does the following:
\begin{itemize}
\item Enumerate  natural numbers $n=1,2,3,\ldots$ one by one.
\item Checks whether $n$ is a proof of $\phi$ using $\gamma_{\mathcal{T}}(\lceil\phi\rceil,n)$.
\end{itemize}
This program finds  a proof of $\phi$ and halts in finite time if $\phi$ is provable in $\mathcal{T}$,
but it does not halt if $\phi$ is not provable in $\mathcal{T}$.

We now consider the program $\xi_{\mathcal{T}}:=\delta_{\mathcal{T}}(\lceil 0=1\rceil )$.
This program halts if and only if there is a derivation of this contradiction from $\mathcal{T}$.
In other words, this program halts if $\text{Con}(\mathcal{T})$ is false,
and does not halt if $\text{Con}(\mathcal{T})$ is true.
G\"odel's second incompleteness theorem says that which of the two alternatives is chosen is independent of $\mathcal{T}$.
In particular, the program $\xi_{\mathcal{ZFC}}$ 
halts if and only if $\mathcal{ZFC}$ is inconsistent,
and whether it does so can neither be proved nor disproved from ${\mathcal{ZFC}}$.
We note that a variant of $\xi_{\mathcal{ZFC}}$, an explicit Turing machine whose halting behavior is independent of $\mathcal{ZFC}$, was explicitly written down in \cite{small},
which only has 7910 internal states.

\subsection{Diophantine equations and Hilbert's 10th problem}
\label{sec:diophantine}
So far, we recalled  that
 there is no algorithm which decides for each Turing machine $\xi$ whether $\xi$ halts in finite steps,
and that there is a specific program $\xi_{\mathcal{ZFC}}$ 
such that whether it halts or not is undecidable in $\mathcal{ZFC}$.
Here we explain, again very briefly, how these facts were  used to answer Hilbert's 10th problem negatively. 
For more details, the readers are referred e.g.~to Davis' review \cite{DavisReview}
or Matiyasevich's textbook \cite{MatiyasevichBook}.

Hilbert's 10th problem is the question of finding an effective method to decide if a given Diophantine equation has a solution or not.
Here, a Diophantine equation is a polynomial equation of many variables with integer coefficients 
over natural numbers,
\begin{equation}
P(x_1,\ldots,x_k)=0, \qquad x_i\in \mathbb{N}.
\end{equation}

A long series of works by Davis, Putnam, Robinson and finally by Matiyasevich 
established that for any Turing machine $\xi$ 
one can construct a Diophantine equation $P_\xi(x_1,\ldots,x_k)=0$ 
such that it has a solution if and only if $\xi$ halts.
This immediately means that there is no algorithm to tell whether a Diophantine equation has a solution or not,
and that there is a specific Diophantine equation $P_{\xi_{\mathcal{ZFC}}}$ such that 
whether it has a solution is independent of $\mathcal{ZFC}$.

Explicitly writing down this equation is a difficult problem.
One method is to use the Diophantine form $P_\upsilon$ of the universal Turing machine $\upsilon$,
which allows us to translate whether a program $\xi$ halts on the input $x$
into whether the Diophantine equation \begin{equation}
P_\upsilon(\lceil \xi\rceil, x; x_1,\ldots,x_k)=0
\end{equation} has a solution $x_1,\ldots,x_k\in \mathbb{N}$ or not.
An example of such a polynomial $P_\upsilon$ was constructed in \cite{UniversalDiophantine},
and has the form \begin{equation}
\begin{aligned}
0&=P_\upsilon(v,x; 
a,b,c,d,e,f,g,h,u,j,k,l,m,n,o,p,q,r,s,t,w,z,u,y,\alpha,\gamma,\eta,\theta,\lambda,\tau,\phi) \\
&= (((zuy)^2+u)^2+y-v)^2
+ (elg^2+\alpha-(b-xy)q^2)^2
+(q-b^{5^{60}})^2\\
&+(\lambda+q^4-1-\lambda b^5)^2
+(\theta+2z-b^5)^2
+(l-u-t\theta)^2
+(e-y-m\theta)^2
+(n-q^{16})^2\\
&+(r-(g+eq^3+lq^5+(2(e-z\lambda)(1+xb^5+g)^4+\lambda b^5 + \lambda b^5 q^4)q^4)(n^2-n)\\
&\qquad -(q^3-bl+l+\theta\lambda q^3+(b^5-2)q^5)(n^2-1))^2\\
&+(p-2ws^2r^2n^2)^2
+(p^2k^2-k^2+1-\tau^2)^2
+(4(c-ksn^2)^2+\eta-k^2)^2\\
&+(k-(r+1+hp-h))^2
+(a-(wn^2+1)rsn^2)^2
+(c-2r-1-\phi)^2\\
&+(d-(bw+ca-2c+4a\gamma-5\gamma))^2
+(d^2-(a^2-1)c^2-1)^2\\
&+(f^2-(a^2-1)i^2c^4-1)^2\\
&+((d+of)^2-((a+f^2(d^2-a))^2-1)(2r+1+jc)^2+1)^2.
\end{aligned}
\label{UniversalDiophantine}
\end{equation}
Then, for a specific choice of $v$ and $x$,
the existence of a solution to this Diophantine equation is equivalent to
the inconsistency of $\mathcal{ZFC}$,
and therefore can neither be proved nor disproved from $\mathcal{ZFC}$.
Finding explicit natural numbers $v,x$ for $\xi_{\mathcal{ZFC}}$ is still a tedious question, however.
Another method is to use the Diophantine equation representing whether a statement $\phi$ is provable in a certain theory $\mathcal{T}$ including elementary arithmetic, which was written down in \cite{CarlMoroz}.

Before proceeding, we point out two minor facts.
First, the question of the existence of a simultaneous solution to a set of Diophantine equations 
$P_1=0$, $P_2=0$, \ldots, $P_\ell=0$ can be reduced to 
the question of the existence of a solution to a single Diophantine equation \begin{equation}
0=P:=(P_1)^2+(P_2)^2+\cdots+(P_\ell)^2.
\end{equation}
In fact, the formula \eqref{UniversalDiophantine} is obtained by using this trick
from a set of equations given in \cite{UniversalDiophantine}.
Second, in the theory of Diophantine equations, it is common to let unknowns $x_1$, \ldots, $x_k$ to vary over natural numbers as we did above,
but the existence of a solution to Diophantine equations over integers is equally undecidable. 
This can be shown using Lagrange's four square theorem, saying that any natural number $x$ can be written as $x=y_1^2+y_2^2+y_3^2+y_4^2$, where $y_i$ are integers.
Indeed, then, the question of the existence of a solution to $P(x_1,\ldots,x_k)=0$ with $x_i\in \mathbb{N}$  can be rewritten as the question of the existence of a solution to \begin{equation}
P(y_{11}^2+y_{12}^2+y_{13}^2+y_{14}^2,
y_{21}^2+y_{22}^2+y_{23}^2+y_{24}^2,
\ldots,
y_{k1}^2+y_{k2}^2+y_{k3}^2+y_{k4}^2)=0
\end{equation} with $y_{ia}\in \mathbb{Z}$,
showing that the question of the existence of a solution over integers is undecidable.
We will use the undecidability in this form later in this article in Sec.~\ref{sec:susy}.

\section{Physics manifestations}
\label{sec:physics}
\subsection{Undecidability in  classical mechanical systems}
\label{sec:classical}

The earliest realizations that the undecidability in mathematics also manifests itself in theoretical physics were given, to the author's knowledge, 
in the context of classical mechanics in the late 1980s to the early 1990s in \cite{GerochHartle,Moore,daCostaDoria}.
Here we would like to review their arguments.

The first reference \cite{GerochHartle} only gave a brief argument on this point, that
one can build a Turing machine within classical mechanics.
This then immediately translates to the fact that it is undecidable if a given outcome is reached in a given classical mechanical system.

The second reference \cite{Moore} gave the argument somewhat more precisely. 
In it, it was pointed out that one can devise a non-linear continuous mapping $f$ of a square onto itself depending on a given Turing machine, 
such that $f(f(f(\cdots((x_0))\cdots)))$ reaches a certain subregion of the square if and only if the Turing machine halts.
Then the author of \cite{Moore} outlined, although not very precisely, that one can design a potential such that a classical point particle moving under its influence will follow the mapping $f$,
thus arguing that the outcome of a given classical mechanical system is undecidable.

The third reference \cite{daCostaDoria} is of a somewhat different nature. 
It uses a mathematical result of \cite{Richardson} saying that
it is undecidable to conclude if a real function given by an explicit mathematical expression composed from elementary functions is actually a zero function or not.
Then, the author of \cite{daCostaDoria}  considers classical systems
whose Hamiltonian is given by explicit mathematical expressions.
Clearly, such systems inherit the undecidability associated to the fact
that it is undecidable to know if a certain term in its Hamiltonian is zero or not.

\subsection{Undecidability of ground state properties of 2d quantum spin systems}
\label{sec:gaplessness}
\def\bC{\mathbb{C}}
We next provide a brief review of the wonderful work \cite{2dNature,2dLong} in which the authors showed that whether the gaplessness  in  the infinite volume limit  of 2d spin systems with the finite-range interaction is undecidable.
This is done by finding a way to construct 2d spin systems from a given Turing machine $\xi$ so that
the property of its ground state depends on whether $\xi$ halts or not.

The main technical result of \cite{2dNature,2dLong} is the following: there exists a nearest-neighbor translationally-invariant  Hamiltonian $H_\xi$ 
on a 2d spin system  with the Hilbert space $\mathcal{H}=\bigotimes_i V_i$,
where the local Hilbert space is $V_i:=\bC^{d_\xi}$,
such that the ground state energy $E_0$ on a lattice of size $L^2$ behaves as 
\begin{equation}
\begin{aligned}
E_0 &\le  -1/2 && \text{for all $L$ if $\xi$ does not halt,}\\
E_0 & \ge +1/2 && \text{for all $L\ge L_\xi$ if $\xi$ halts,}
\end{aligned}
\end{equation}
where $L_\xi$ is a constant related to the running time of the program $\xi$.

The papers \cite{2dNature,2dLong} actually describe two different versions of this result.
One is their Lemma 8, where the dimension $d_\xi$ of the local Hilbert space is at most proportional to the number of states of the Turing machine $\xi$, but with the bonus that the ground state in the non-halting case is a gapped, product state.
Another is their Corollary 54, where $d_\xi$ is now independent of $\xi$, but without the guarantee of the nature of the ground state in the halting case.
We use the first version in the rest of the section.

The construction of this Hamiltonian $H_\xi$ requires real ingenuity, which the interested readers  can marvel at by going over their detailed description in \cite{2dLong},
and which the author of this present article dare not to provide. 
The remaining task is to translate this undecidability to arbitrary properties of the ground state.
For this purpose, we take two auxiliary 2d nearest-neighbor spin systems whose behavior we do understand,
$H_W$ acting on $\bigotimes_i W_i$  with $W_i=\bC^a$,
and 
$H_Z$ acting on $\bigotimes_i Z_i$  with $Z_i=\bC^b$,
both normalized so that the ground state has zero energy.
Let us write the nearest-neighbor interactions of $H_\xi$, $H_W$ and $H_Z$ as 
$h_\xi^{(i,j)}$, $h_W^{(i,j)}$ and $h_Z^{(i,j)}$.
We now fuse three systems as follows.
We take the local Hilbert space per site of the combined system to be $U_i=V_i\otimes W_i\oplus Z_i$,
and we take the new nearest-neighbor Hamiltonian to be given by
\begin{equation}
H_\text{combined} = 
\sum_{i,j} (h_\xi^{(i,j)} + h_W^{(i,j)} + h_Z^{(i,j)})
+ \lambda \sum_{i,j} P_{V_i\otimes W_i} P_{Z_{i}}.
\end{equation}
Here, 
the term $h_\xi^{(i,j)} + h_W^{(i,j)}$ now acts on $V_i \otimes W_i \otimes V_{j}\otimes W_{j} \subset U_i\otimes U_{j}$ extended by zero outside by abuse of notations,
the term $h_Z^{(i,j)}$ now acts on $Z_i \otimes Z_{j} \subset U_i\otimes U_{j}$ similarly extended by zero,
$\lambda$ is a large positive constant, and 
$P_{V_i\otimes W_i}$ and $P_{Z_i}$ are projectors to the respective components. 

Let us study the ground state of the combined system.
The term proportional to $\lambda$ gives a huge penalty to wavefunctions supported both on the $V\otimes W$ part and on the $Z$ part. 
Therefore, the ground state comes either from wavefunctions on the $V\otimes W$ part only,
or those on the $Z$ part only.
The ground state on the $Z$ part is that of $H_Z$ and the energy is $E_0^{(Z)}=0$.
The ground state on the $V\otimes W$ depends on whether $\xi$ halts or not.
If it halts, the energy $E_0^{(V\otimes W)}\ge +1/2$. 
If it does not halt, $E_0^{(V\otimes W)}\le -1/2$,
and the ground state has all the features of that of $H_W$,
since the ground state of $H_\xi$ in this case is guaranteed to be a gapped product state.
Therefore, the ground state of the combined system is that of $H_W$ when $\xi$ does not halt,
and is that of $H_Z$ when $\xi$ does halt.

This means that essentially any property of the ground state of 2d spin systems is undecidable.
For example, by taking $H_W$ to be gapped and $H_Z$ to be gapless, 
one finds that the gaplessness of the 2d spin systems is undecidable.
Similarly, by taking $H_W$ and $H_Z$ to be different symmetry protected topological (SPT) phases   
protected by a symmetry group $G$, 
one finds that determining the SPT phase given a $G$-symmetric gapped  system is also an undecidable problem.

\subsection{Undecidability of supersymmetry breaking}
\label{sec:susy}

Let us finally come to a case of undecidability in quantum field theory.
This subsection is the only new result in this article, and yet 
it is an immediate consequence of the basic materials reviewed above.\footnote{%
This was the reason why the author of the present article felt it impossible to make this document public unless as a bedtime reading as a preprint on April Fools' day.
The author thanks an editor of this journal for the invitation to submit and the referees for consideration to publish.}

Our claim is the following.
First, there is no algorithm which decides, for each 2d \Nequals{(2,2)}  supersymmetric Lagrangian theory given to it, whether it breaks supersymmetry.
Second, there is a specific 2d \Nequals{(2,2)} supersymmetric Lagrangian theory
which breaks supersymmetry if and only if $\mathcal{ZFC}$ is consistent,
which can never be proved nor disproved from $\mathcal{ZFC}$.

We show this by resorting to the negative solution to Hilbert's 10th problem\footnote{%
Possible relevance of the negative solution to Hilbert's 10th problem to the issues surrounding string landscape was previously studied in \cite{Cvetic:2010ky,Halverson:2019vmd}.
It is to be noted that the undecidability of the quantum control in \cite{Control} also reduces the issue to the undecidability of the Diophantine problem.
} we reviewed in Sec.~\ref{sec:diophantine}.
Take a Diophantine equation \begin{equation}
P_\xi(x_1,\ldots,x_k)=0 \label{D}
\end{equation} over $\mathbb{Z}$ corresponding to a Turing machine $\xi$.
We now consider a Wess-Zumino model with $2k+1$ chiral superfields
$Y$,  $Z_1$, \ldots, $Z_k$
and $X_1$, \ldots, $X_k$, with the superpotential \begin{equation}
W_\xi= 
YP_\xi(X_1,\ldots,X_k)^2 + 
\sum_a Z_a (\sin 2\pi i X_a)^2 .
\end{equation}

Let us now look for supersymmetric vacua of this model.
The condition  $\partial W_\xi/\partial Z_a=0$ imposes the condition $X_a\in \mathbb{Z}$.
Then the  condition  $\partial W_\xi/\partial Y=0$ imposes  the condition $P_\xi=0$.
When these two conditions are met, the remaining F-term conditions $\partial W_\xi/\partial X_a=0$ are  automatically satisfied, 
without restricting the vacuum expectation values of $Z_a$ and $Y$.
We found that this model breaks supersymmetry if and only if the Diophantine equation $P_\xi$ does not have any solution over integers,
i.e.~if and only if  the Turing machine $\xi$ does not halt.

Now, the undecidability of the halting of Turing machines immediately means that 
there is no uniform algorithm to decide if a given 2d \Nequals{(2,2)}-supersymmetric Wess-Zumino model is supersymmetric or not.
Furthermore, by using the Turing machine $\xi_{\mathcal{ZFC}}$ which looks for a contradiction of $\mathcal{ZFC}$,
we can construct a Wess-Zumino model with the superpotential $W_{\xi_{\mathcal{ZFC}}}$ 
which breaks supersymmetry 
if and only if $\mathcal{ZFC}$ is consistent,
which can neither be proved nor be disproved by means of $\mathcal{ZFC}$ itself.

\section{Discussions}
\label{sec:discussions}
In this paper, we reviewed basic mathematical concepts surrounding undecidability in an informal manner.
We then proceeded to recall the mechanisms how such mathematical decidability was translated to the realm of theoretical physics,
first in \cite{GerochHartle,Moore,daCostaDoria} in the case of classical mechanics,
and second in \cite{2dNature,2dLong} and the subsequent works in the case of quantum mechanical systems.
Finally we presented a minor but new result, that the existence of supersymmetric vacua in \Nequals{(2,2)}-supersymmetric  Wess-Zumino models is similarly undecidable.

As already mentioned in the introduction and explained in more detail in the main text,
the strategy is always the same.
Namely, to show that a particular property of a certain class of systems in theoretical physics is undecidable,
we only have to devise a way to encode arbitrary Turing machines into the class of systems under consideration. 
It can be done directly as in \cite{2dNature,2dLong}, or indirectly by first translating to the Diophantine problem as in this work or in \cite{Control}, 
or to other mathematical formulations as in \cite{daCostaDoria} which used \cite{Richardson}.

This makes it clear that the easier the construction of undecidable problems become,
the broader the class of physical systems one wants to consider.
For example, in the paper \cite{GerochHartle} we reviewed, the entire classical physics was considered, in which clearly a computer can be implemented.
Therefore it was almost immediate that their question at hand was undecidable,
and there was no element of surprise in it.

In contrast, in the case of \cite{2dNature,2dLong}, 
the class of systems they considered, namely quantum spin systems
with finite-dimensional Hilbert space at each site
with translation-invariant and finite-range interactions,
was felt to be narrow enough, at least to a large percentage of the researchers in that field, in the understanding of the author of the present article.
This explains the surprise with which the results of \cite{2dNature,2dLong} were received,
and also the flurry of subsequent works generalizing and extending them.

How should we judge the only new result of this paper, presented in Sec.~\ref{sec:susy}, in view of these considerations?
It should be pointed out that the superpotential of \Nequals{(2,2)} supersymmetric quantum field theory in two dimensions can be an arbitrary holomorphic function.
This was the crucial property which allowed us to encode Diophantine equations easily into this class of systems. 
Therefore it would not be very fair to say that the result is of great importance, 
except possibly to the author of this article himself, since it allowed him to dispel 
his childlike trust in the collective intellect of theoretical physicists in solving questions in theoretical physics. 

For a possible direction of future research, 
we can consider, for example, if we can encode Turing machines
into the properties of renormalizable  quantum field theories in four dimensions,
supersymmetric or otherwise.
This is the standard class of quantum field theories high-energy physicists habitually work in, 
and include the Standard Model of particle physics and their various extensions.
The main difficulty, compared to the two-dimensional case considered in this paper, 
is that the (super)potential of this class of theories only allows
polynomials of degrees three (in the case of superpotentials of supersymmetric theories) or four (in the case of potentials of non-supersymmetric theories).
Therefore the class is much more restricted, 
and the encoding of Turing machines would accordingly be more difficult.

The author of the present article does not know if this would be possible or not.
Would we learn anything new, if it would turn out to be possible?
That is debatable, as the many works e.g.~\cite{Control,thermalNature,PhaseDiagram,RG}, which followed the seminal paper \cite{2dNature}, only show that the 
same technique of embedding of Turing machines works in various types of properties of
various classes of systems, 
and the encoding into four-dimensional quantum field theories merely provides one more.
That said, it would surely be intellectually interesting, at least to the author of the present article.


\section*{Acknowledgments}
The author thanks Beni Yoshida for his tweet\footnote{\url{https://twitter.com/rougeteaviolet/status/1395922985614299137}} which introduced me to the wonderful paper \cite{Cubitt2nd}.
He also thanks helpful discussions with Justin Kaidi and Kantaro Ohmori.
The content of this paper was presented first in a lunch talk at IPMU in the autumn, 2021;
the author thanks  the audience members for the lively and stimulating discussions there. 
The author is supported in part  by WPI Initiative, MEXT, Japan at Kavli IPMU, the University of Tokyo.

This article originally appeared on the arXiv preprint server on April Fool's day of 2022, in the hope of  providing some intellectual entertainment in that dark year.
The author thanks the editor of this journal for suggesting him to submit the preprint, which he did not originally have any intention to publish.
The editor's suggestion to submit led to a fruitful refereeing process, allowing the author to improve the manuscript greatly.
For this the author would also like to thank the two referees.

This article was written down during a week in January 2022 when 
he and his wife  needed to take care of the children by themselves
because the kindergarten was closed due to the omicron variant.
One day, while preparing the table for the supper, he told his wife that 
he finally finished an article for April Fools' day, to which his wife exclaimed:
``We can work only half as usual, because one of us needs to take care of the kids
while the other is working,
and then \emph{you} were using that precious time to write an \emph{article for April Fools' day}?!?''
It took the author a couple of minutes to explain that the article was not just to make stupid jokes 
but to make somewhat unusual philosophical takes on theoretical physics. 
The author is not sure if that explanation successfully convinced his wife.

\def\arxivfont{\rm}
\bibliographystyle{ytamsalpha}

\bibliography{ref}

\end{document}